# Multiparameter spectral analysis for aeroelastic instability problems

Arion Pons[1,*] and Stefanie Gutschmidt[2]

[1]Department of Engineering, University of Cambridge, Cambridge CB2 1PZ, UK.
[2]Department of Mechanical Engineering, University of Canterbury, Christchurch 8140, New Zealand.

**Key words**      multiparameter eigenvalue problems; instability; aeroelasticity.

**AMS classification**      97M50; 65F15; 15A18; 15A69

This paper presents a novel application of multiparameter spectral theory to the study of structural stability, with particular emphasis on aeroelastic flutter. Methods of multiparameter analysis allow the development of new solution algorithms for aeroelastic flutter problems; most significantly, a direct solver for polynomial problems of arbitrary order and size, something which has not before been achieved. Two major variants of this direct solver are presented, and their computational characteristics are compared. Both are effective for smaller problems arising in reduced-order modelling and preliminary design optimization. Extensions and improvements to this new conceptual framework and solution method are then discussed.

## 1. Introduction

Predicting and controlling aeroelastic instability forms a major part of the discipline of aeroelasticity. Many different physical systems show flutter instability, and many models exist to describe them. In a linear system, or the linearization of a nonlinear system, the onset of flutter or divergence can be formulated in the well-known stability criterion:

$$\text{Im}(\chi) > 0 \text{ for stability} \qquad (1)$$

where $\chi$ are the time-eigenvalues of the system according to the Fourier transform $q(t) = \hat{q}e^{i\chi t}$ for the system coordinate $q$ [1]. These eigenvalues may be nondimensionalised. Flutter occurs when the system parameters are such that the system is on the stability boundary, $\text{Im}(\chi_f) = 0$. The flutter point may then be described as a tuple which includes the set of system parameters (particularly, an airspeed parameter) and the modal frequency of instability, $\chi_f$. Typically only one or two flutter or divergence points are of industrial relevance.

Eq. 1 is not however the only criterion that can be used to characterize instability. This leads us into the study of aeroelastic methods. Apart from the four established methods – the *p-method*, *classical flutter analysis*, the *k-method* (or *U-g method*, or *V-g method*) and the *p-k method*; all of which are detailed and discussed in a number of reference works [1–3] – recent years have seen a proliferation of new aeroelastic methods. Several authors have refined the existing established methods for particular scenarios or applications [4–6]. The application of concepts from robust control theory have yielded a series of methods, including the $\mu$-method by Lind and Brenner [7,8], the $\mu$-k method by Borglund [9–12], and others [13–15]. The prime advantage of these $\mu$-type methods is that they facilitate the propagation of uncertainty distributions through the system, allowing a worst-case flutter speed estimate to be made in a system with high uncertainty. Other developments have come from other fields: Afolabi [16,17] characterized coupled-mode flutter as a loss of eigenvector orthogonality, using methods from catastrophe theory. Irani and Sazesh [18] used stochastic methods, while Gu et al. [19] devised a genetic algorithm, and a number of authors [20–23] have applied neural networks to the detection of flutter points.

---

* Corresponding author, adp53@cam.ac.uk, tel. +44 755 366 3296

It is in the context of these developments that we propose our method of analyzing flutter problems. The central methodological contribution of this paper is the concept that the solution of an aeroelastic system for its flutter points is nothing other than a multiparameter eigenvalue problem. We will show the simple link between the aeroelastic stability problem and multiparameter spectral theory, and how this allows for direct solution of a variety of flutter problems. The purpose of this paper is to detail this link and to explore the mechanisms by which this direct solution may be accomplished. For this purpose we will apply our analysis to simple example problems – however, the method does extend to problems that are significantly more complex; these are considered in Pons and Gutschmidt [24, 25].

## 2. Aeroelastic flutter as a multiparameter eigenvalue problem (MEP)

Consider a linear finite-dimensional system with eigenvector $\mathbf{x} \in \mathbb{C}^n$ and arbitrary continuous dependence on both an eigenvalue parameter $\chi \in \mathbb{C}$, and another structural or environmental parameter $p \in \mathbb{R}$:

$$A(\chi, p)\mathbf{x} = \mathbf{0} \qquad (2)$$

where $A \in \mathbb{C}^{n \times n}$. Any complex structural parameter can of course be split into two real parameters. The stability problem for this system (with respect to parameter $p$) is to find $p$ such that an eigenvalue of the problem ($\chi$) has zero imaginary part. This point is the 'stability boundary': for a system with multiple structural parameters, the stability boundary may be a line or other higher-dimensional surface. We then note that the condition $\text{Im}(\chi) = 0$ is equivalent to modifying the original definition of the problem such that $\chi \in \mathbb{R}$ and not $\chi \in \mathbb{C}$. Such a manoeuvre does not seem to be immediately useful: under $\chi \in \mathbb{R}$, a solution to Eq. 2 only exists on the stability boundary, and nowhere else. In order to develop, for example, iterative methods for flutter point calculation, we need to be able to define some form of solution in the subcritical and supercritical areas (above and below the stability boundary, respectively). There is an easy way of doing this. Following [26], we take the complex conjugate of Eq. 2 as another equation:

$$\begin{aligned} A(\chi, p)\mathbf{x} &= \mathbf{0}, \\ \overline{A}(\chi, p)\overline{\mathbf{x}} &= \mathbf{0}. \end{aligned} \qquad (3)$$

As $p \in \mathbb{R}$ and $\chi \in \mathbb{R}$ are unaffected by the conjugation, this operation enforces these conditions. This procedure has been utilized before in the analysis of delay differential equations [26], and (in a limited form) in the context of Hopf bifurcation prediction [27], but has never been applied to aeroelastic or other structural stability problems. Equation 3 is nothing other than a multiparameter eigenvalue problem (MEP): an eigenvalue problem in which the eigenvalue point is not simply defined by a scalar and an eigenvector, but by an $n$-tuple and an eigenvector. A number of methods of analysis have been developed for such problems, and in this paper we will explore some of these. However, as the methods that are available depend strongly on the structure of matrix A, we will first define a system to work with.

## 3. An Example Section model

### 3.1. Formulation

Consider first the simple section model shown in Figure 1. This model has two degrees of freedom: plunge $h$ and twist $\theta$. The governing equations for this model are easy to derive; they are:

$$\begin{aligned} m\ddot{h} + d_h\dot{h} + k_h h - mx_\theta\ddot{\theta} &= -L(t), \\ I_P\ddot{\theta} + d_\theta\dot{\theta} + k_\theta\theta - mx_\theta\ddot{h} &= M(t), \end{aligned} \quad (4)$$

where $m$ and $I_P$ are the section mass and polar moment of inertia, $k_h$ and $k_\theta$ are the section plunge and twist stiffnesses, $d_h$ and $d_\theta$ are the section plunge and twist damping coefficients, and $x_\theta$ is the section's static imbalance – defined as the distance along the $x$-axis from the pivot point to the centre of mass. Taking the Fourier transform, $[h(t), \theta(t)] = [\hat{h}, \hat{\theta}]e^{\iota\chi t}$, of this model, we obtain:

$$\begin{aligned} (-m\chi^2 + \iota d_h\chi + k_h)\hat{h} + mx_\theta\chi^2\hat{\theta} &= L(\chi, \hat{h}, \hat{\theta}), \\ mx_\theta\chi^2\hat{h} + (-I_P\chi^2 + \iota d_\theta\chi + k_\theta)\hat{\theta} &= M(\chi, \hat{h}, \hat{\theta}). \end{aligned} \quad (5)$$

To model the aerodynamic loads in the frequency domain we use Theodorsen's unsteady aerodynamic theory [3]:

$$L = -\chi^2(L_h\hat{h} + L_\theta\hat{\theta}), \quad M = \chi^2(M_h\hat{h} + M_\theta\hat{\theta}). \quad (6)$$

The aerodynamic coefficients $\{L_h, L_\theta, M_h, M_\theta\}$ are complex functions of $\kappa$ – the reduced frequency; an aerodynamic parameter related to the airspeed ($U$) by $\kappa = b\chi/U$. Other structural and environmental parameters involved are the air density ($\rho$), the semichord ($b$) and the distance along the $x$-axis from the midchord to the pivot point, as a fraction of the semichord ($a$). See Pons [24] or Hodges and Pierce [3] for details. We assume that the flow over the airfoil is quasisteady: that is, that Theodorsen's function takes a value of 1 universally [1,3]. We will deal with general Theodorsen aerodynamics in a later paper, as this requires iterative or approximate multiparameter solution methods which we will not cover here. We also assume without loss of generality a lift-angle of attack coefficient of $C_{L\alpha} = 2\pi$, as per thin-airfoil theory [28].

It is then customary to nondimensionalise Eq. 5. Further details of this are given in Pons [24]. The final result is a flutter problem of the form

$$\left(\left(M_0 + G_0 + G_1\frac{1}{\kappa} + G_2\frac{1}{\kappa^2}\right)\chi^2 - D_0\chi - K_0\right)\mathbf{x} = \mathbf{0}. \quad (7)$$

with dimensionless parameters defined as in Table 1 and the nomenclature, and the matrix coefficients

$$G_0 = \frac{1}{\mu}\begin{bmatrix} 1 & a \\ a & \left(\frac{1}{8} + a^2\right) \end{bmatrix},$$

$$G_1 = \frac{1}{\mu}\begin{bmatrix} -2\iota & 2\iota(1-a) \\ -\iota(1+2a) & \iota a(1-2a) \end{bmatrix}, \quad (8)$$

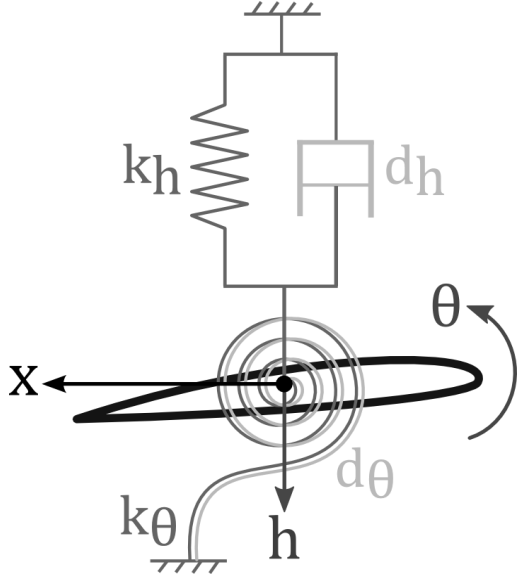

**Table 1**: Dimensionless parameter values for the section model

| Parameter | Value |
|---|---|
| mass ratio – $\mu$ | 20 |
| radius of gyration – $r$ | 0.4899 |
| bending nat. freq. – $\omega_h$ | 0.5642 rad/s |
| torsional nat. freq. – $\omega_\theta$ | 1.4105 rad/s |
| bending damping – $\zeta_h$ | 1.4105 % |
| torsional damping – $\zeta_\theta$ | 2.3508 % |
| static imbalance – $r_\theta$ | −0.1 |
| pivot point location – $a$ | −0.2 |

**Figure 1:** Diagram of section model

$$G_2 = \frac{1}{\mu}\begin{bmatrix} 0 & 2 \\ 0 & 1+2a \end{bmatrix}, \quad M_0 = \begin{bmatrix} 1 & -r_\theta \\ -r_\theta & r^2 \end{bmatrix},$$

$$D_0 = \begin{bmatrix} 2\iota\zeta_h\omega_h & 0 \\ 0 & 2\iota r^2 \zeta_\theta \omega_\theta \end{bmatrix}, \quad K_0 = \begin{bmatrix} \omega_h^2 & 0 \\ 0 & r^2 \omega_\theta^2 \end{bmatrix}, \tag{8}$$

Note that this nondimensionalisation is a convenience and not necessary prerequisite for using multiparameter solution methods. However something that will be of great use is to rearrange Eq. 7 into two polynomial forms by defining new eigenvalue parameters: $\Upsilon = U/b$, $\tau = 1/\kappa$, and $\lambda = 1/\chi$. These two forms are then

$$\big((M_0 + G_0)\chi^2 + G_1 \Upsilon \chi + G_2 \Upsilon^2 - D_0 \chi - K_0\big)x = 0, \tag{9}$$

which we term the $\Upsilon$-$\chi$ form, and

$$\big((M_0 + G_0) + G_1 \tau + G_2 \tau^2 - D_0 \lambda - K_0 \lambda^2\big)x = 0, \tag{10}$$

which we term the $\tau$-$\lambda$ form. Both are preferable to the $k$-$\chi$ form, Eq. 7. We should note at this point that, although we have here derived these two forms for a small 2-DOF section model, these forms arise in many other models; often with significantly larger matrices. The solution methods we develop for this form will thus be applicable to a wide variety of problems.

### 3.2. Undamped system

In many aeroelastic systems structural damping is negligible, in which case $D_0 = 0$. While this may seem to be a trivial case of the preceeding systems, the omission of the structural damping term does allow us to change the structure of the system, leading to a faster computation time (as we will show later). In the case of Eq. 10, we have

$$\big((M_0 + G_0) + G_1 \tau + G_2 \tau^2 - K_0 \Lambda\big)x = 0, \tag{11}$$

where $\Lambda = \lambda^2$. This system is now linear in $\Lambda$, whereas it was previously quadratic in $\lambda$. Note that the $\tau$-$\chi$ form is the only polynomial form presented which allows us to make one parameter linear in this situation.

### 3.3. Transformation into MEPs

The three polynomial systems that we have introduced in this section (Eq. 9, 10 and 11) are not yet fully constrained MEPs, as they only consist of one equation. To constrain the system we employ the method introduced in Section 2 – the addition of an equation representing the conjugate of the initial equation. We obtain:

$$\begin{aligned}\left((M_0 + G_0) + G_1\tau + G_2\tau^2 - K_0\Lambda\right)\mathbf{x} &= \mathbf{0}, \\ \left((\overline{M}_0 + \overline{G}_0) + \overline{G}_1\tau + \overline{G}_2\tau^2 - \overline{K}_0\Lambda\right)\overline{\mathbf{x}} &= \mathbf{0},\end{aligned} \quad (12)$$

$$\begin{aligned}\left((M_0 + G_0) + G_1\tau + G_2\tau^2 - D_0\lambda - K_0\lambda^2\right)\mathbf{x} &= \mathbf{0}, \\ \left((\overline{M}_0 + \overline{G}_0) + \overline{G}_1\tau + \overline{G}_2\tau^2 - \overline{D}_0\lambda - \overline{K}_0\lambda^2\right)\overline{\mathbf{x}} &= \mathbf{0},\end{aligned} \quad (13)$$

and

$$\begin{aligned}\left((M_0 + G_0)\chi^2 + G_1\Upsilon\chi + G_2\Upsilon^2 - D_0\chi - K_0\right)\mathbf{x} &= \mathbf{0}, \\ \left((\overline{M}_0 + \overline{G}_0)\chi^2 + \overline{G}_1\Upsilon\chi + \overline{G}_2\Upsilon^2 - \overline{D}_0\chi - \overline{K}_0\right)\overline{\mathbf{x}} &= \mathbf{0}.\end{aligned} \quad (14)$$

These systems are all polynomial multiparameter eigenvalue problems [29]. A number of solution methods are known for such systems.

## 4. Direct solution via linearization

### 4.1. Linearization

Any polynomial MEP can be made into a linear one (consisting of a zeroth-order term and first-order terms in each eigenvalue) via the process of linearization [30]. This process bears resemblance to the linearization of single-parameter polynomial eigenvalue problems; a process which is well-known [31]. Multiparameter linearization is relevant because a number of direct solution methods exist for linear MEPs. Consider first Eq. 12. This equation contains only one quadratic variable ($\tau$), as $\Lambda$ is already linear. To linearize this system, we define a new eigenvector which contains a factor of $\tau$: $\mathbf{q} = [\mathbf{x}; \tau\mathbf{x}]$. By expanding the system's coefficient matrices and using this factor of $\tau$ in the eigenvector to reduce the order of the quadratic term, we obtain a linear problem of double the size:

$$\begin{aligned}\left(\begin{bmatrix}M_0 + G_0 & 0 \\ 0 & -I_n\end{bmatrix} + \begin{bmatrix}-K_0 & 0 \\ 0 & 0\end{bmatrix}\Lambda + \begin{bmatrix}G_1 & G_2 \\ I_n & 0\end{bmatrix}\tau\right)\mathbf{q} &= \mathbf{0} \\ \left(\begin{bmatrix}\overline{M}_0 + \overline{G}_0 & 0 \\ 0 & -I_n\end{bmatrix} + \begin{bmatrix}-\overline{K}_0 & 0 \\ 0 & 0\end{bmatrix}\Lambda + \begin{bmatrix}\overline{G}_1 & \overline{G}_2 \\ I_n & 0\end{bmatrix}\tau\right)\overline{\mathbf{q}} &= \mathbf{0}\end{aligned} \quad (15)$$

The upper rows of Eq. 15 represents Eq. 12 directly, and the lower row represents the identity $I_n(\tau\mathbf{x}) = \tau I_n\mathbf{x}$. Note that other linearizations of similar form are possible. However, irrespective of the exact linearization used, the resulting system will be of the form:

$$\begin{aligned}(A + B\Lambda + C\tau)\mathbf{q} &= \mathbf{0}, \\ (\overline{A} + \overline{B}\Lambda + \overline{C}\tau)\overline{\mathbf{q}} &= \mathbf{0}.\end{aligned} \quad (16)$$

Note that it is not actually necessary to linearize the second (i.e. conjugate) equation of the aeroelastic system, because the linearized conjugate equation will be the conjugate of the linearized first equation. This is a property of this method of linearization.

This same linearization process can be applied to any quadratic MEP. Any problem of the form

$$(A + B\lambda + C\tau + D\lambda\tau + E\lambda^2 + F\tau^2)\mathbf{x} = \mathbf{0}, \quad (17)$$

can be linearized as

$$\left(\begin{bmatrix} A & B & C \\ 0 & -I_n & 0 \\ 0 & 0 & -I_n \end{bmatrix} + \begin{bmatrix} 0 & D & E \\ I_n & 0 & 0 \\ 0 & 0 & 0 \end{bmatrix}\lambda + \begin{bmatrix} 0 & 0 & F \\ 0 & 0 & 0 \\ I_n & 0 & 0 \end{bmatrix}\tau\right)\begin{bmatrix} \mathbf{x} \\ \lambda\mathbf{x} \\ \tau\mathbf{x} \end{bmatrix} = \mathbf{0}. \quad (18)$$

We have

$$\begin{aligned} A &= M_0 + G_0 & D &= -K_0 \\ B &= G_1 & E &= 0 \\ C &= -D_0 & F &= G_2 \end{aligned} \quad (19)$$

for Eq. 13, and changing $\lambda \to \chi$ and $\tau \to \Upsilon$,

$$\begin{aligned} A &= -K_0 & D &= M_0 + G_0 \\ B &= -D_0 & E &= G_1 \\ C &= 0 & F &= G_2 \end{aligned} \quad (20)$$

for Eq. 14. These linearised systems have matrix coefficients triple the size of the original problem.

### 4.2. Direct solution

Consider Eq. 16. Post-multiplying the first equation in this system by $\bar{C}\mathbf{y}$ and premultiplying the second by $C\mathbf{x}$, we have

$$\begin{aligned} (A + B\Lambda + C\tau)\mathbf{x} \otimes (\bar{C}\mathbf{y}) &= 0, \\ (C\mathbf{x}) \otimes (\bar{A} + \bar{B}\Lambda + \bar{C}\tau)\mathbf{y} &= 0. \end{aligned} \quad (21)$$

These two equations are both equal to zero so we may equate them. After cancelling the terms in $\tau$, the expression can be manipulated into:

$$\Delta_1 \mathbf{z} = \Lambda \Delta_0 \mathbf{z} \quad (22)$$

and an enlarged eigenvector $\mathbf{z} = \mathbf{x} \otimes \mathbf{y}$ and the operator determinants

$$\begin{aligned} \Delta_0 &= B \otimes \bar{C} - C \otimes \bar{B} \\ \Delta_1 &= C \otimes \bar{A} - A \otimes \bar{C} \\ \Delta_2 &= A \otimes \bar{B} - B \otimes \bar{A}. \end{aligned} \quad (23)$$

Eq. 22 is a generalized eigenvalue problem (GEP), in the single parameter $\lambda$. Solvers for the generalized eigenvalue problem are very widely available. If the linear system has square coefficients of size $m$ then the operator determinants are of size $m^2$.

The operator determinants can also be used to define a second GEP in $\tau$. By multiplying the first and second equations of Eq. 16 by $\bar{B}\mathbf{y}$ and $B\mathbf{x}$ respectively, we can also show that:

$$\Delta_2 \mathbf{z} = \tau \Delta_0 \mathbf{z}. \quad (24)$$

However, it is only necessary to solve one of Eq. 22 or Eq. 24: once one has been solved, then its solutions can be substituted back into the original polynomial system, which yields another smaller GEP. Alternatively, if the eigenvalue is simple then it may be computed more cheaper via Rayleigh quotients in $\mathbf{z}$ or (decomposing $\mathbf{z} = \mathbf{x} \otimes \mathbf{y}$), $\mathbf{x}$ and $\mathbf{y}$. The system is thus completely solved for its flutter points. This direct solution method is known in mathematical literature as the *operator determinant method*. Its computational complexity is $\mathcal{O}(n^6)$, irrespective of whether $n$ is the size of the linear coefficients (A, B, etc.) or the polynomial coefficients ($M_0$, $G_0$, etc.) [30,32,33]. This large complexity arises from solving the generalized eigenvalue problem, an $\mathcal{O}(m^3)$ process by the QZ algorithm [34], with operator determinants of size $m = \mathcal{O}(n^2)$. The operator determinant method has not previously been used in aeroelasticity, and has only rarely seen engineering application in the study of dynamic model updating [35,36].

### 4.3. Singularity

One important caveat of the operator determinant approach is that the matrix $\Delta_0$ must not be singular. If it is, then the eigenvalues of the original polynomial system (e.g. Eq. 12-14) will not generally coincide with those of the GEPs of the linearized problem (Eq. 22 and 24) [30,37,38]. A linear MEP with singular $\Delta_0$ is said to be singular MEP. A large proportion of the linear flutter problems that arise in the study of aircraft aeroelasticity are singular. This is largely because the linearization of polynomial problems tends to generate singular linear problems, even if the original polynomial problem has all its matrix coefficients at full rank – compare Eq. 18. Some smaller linearizations are not necessarily singular: for example Eq. 15. However, here we find that (for our aerodynamic model) the coefficient matrix $G_2$ is not full rank and so the linearized problem is singular anyway. In many circumstances we will be dealing with a singular problem, and for a long time the lack of a working solver for such singular problems has been a major obstacle to the application of multiparameter methods to real-world problems.

However, recently a solution has been proposed. Muhič and Plestenjak [29] proved that the eigenvalues of a polynomial system are equivalent to the finite regular eigenvalues of the pair of singular operator determinant GEPs constructed via linearization. The finite regular eigenvalues of a linear two-parameter system are the pairs $(\lambda, \mu)$ such that $i \in \{1,2\}$ [37]:

$$\text{rank}(A_i + B_i\lambda + C_i\mu) < \max_{(s,t)\in\mathbb{C}^2} \text{rank}(A_i + B_i s + C_i t), \qquad (25)$$

that is, the finite regular eigenvalues are the set of points that cause the singular problem to have its maximum rank, even though this is not full rank. On the basis of this proof, Muhič and Plestenjak [29] devised a set of algorithms which would extract the common regular part of the singular matrix pencils (i.e. matrix-valued functions polynomial in a variable $\lambda$) $\Delta_{1s} - \lambda\Delta_{0s}$ and $\Delta_{2s} - \lambda\Delta_{0s}$. This common regular part is represented by two smaller nonsingular matrix pencils ($\Delta_1 - \lambda\Delta_0$ and $\Delta_2 - \lambda\Delta_0$), the eigenvalues of which are the finite regular eigenvalues of the singular problem and thus the eigenvalues of the polynomial problem. In practical terms, this can be seen as a compression of the singular operator determinant matrices into smaller full-rank matrices. The operator determinant method, as presented in Section 4.2, can be applied to these compressed pencils. The algorithms involved in the extraction of the common regular part are presented in [29] and published also in MATLAB code [39]. We will not detail them here as they are complex.

## 5. Direct solution via Quasi-linearization

### 5.1. Quasi-linearization

Hochstenbach et al. [30] recently presented another method of linearization for polynomial MEPs. Instead of increasing the size of the coefficient matrices, this method of linearization increases the number of parameters and equations in the system. To differentiate it from strict linearization, Hochstenbach et al. [30] term this new method *quasi-linearization*. The process is as follows. Considering again Eq. 12, we define a new eigenvalue parameter $\alpha = \tau^2$. The equations then become

$$((M_0 + G_0) + G_1\tau + G_2\alpha - K_0\Lambda)x = 0,$$
$$((\overline{M}_0 + \overline{G}_0) + \overline{G}_1\tau + \overline{G}_2\alpha - \overline{K}_0\Lambda)\overline{x} = 0. \qquad (26)$$

This is now a linear three-parameter eigenvalue problem, but with only two equations. We need a third equation to constrain the system. We have the relation $\alpha - \tau^2 = 0$, but this is nonlinear. However, noticing that we can write this relation as

$$\det\left(\begin{bmatrix} \alpha & \tau \\ \tau & 1 \end{bmatrix}\right) = 0 \qquad (27)$$

we can recast it as a new MEP to add to Eq. 26 (using an arbitrary eigenvector **y**):

$$((M_0 + G_0) + G_1\tau + G_2\alpha - K_0\Lambda)x = 0,$$
$$((\overline{M}_0 + \overline{G}_0) + \overline{G}_1\tau + \overline{G}_2\alpha - \overline{K}_0\Lambda)\overline{x} = 0, \qquad (28)$$
$$\left(\begin{bmatrix} 0 & 0 \\ 0 & 1 \end{bmatrix} + \begin{bmatrix} 0 & 1 \\ 1 & 0 \end{bmatrix}\tau + \begin{bmatrix} 1 & 0 \\ 0 & 0 \end{bmatrix}\alpha\right)y = 0.$$

In a similar way, we can linearize Eq. 13 with the definitions $\alpha = \tau^2$ and $\beta = \lambda^2$:

$$((M_0 + G_0) + G_1\tau + G_2\alpha - D_0\lambda - K_0\beta)x = 0,$$
$$((\overline{M}_0 + \overline{G}_0) + \overline{G}_1\tau + \overline{G}_2\alpha - \overline{D}_0\lambda - \overline{K}_0\beta)\overline{x} = 0,$$
$$\left(\begin{bmatrix} 0 & 0 \\ 0 & 1 \end{bmatrix} + \begin{bmatrix} 0 & 1 \\ 1 & 0 \end{bmatrix}\tau + \begin{bmatrix} 1 & 0 \\ 0 & 0 \end{bmatrix}\alpha\right)y_1 = 0, \qquad (29)$$
$$\left(\begin{bmatrix} 0 & 0 \\ 0 & 1 \end{bmatrix} + \begin{bmatrix} 0 & 1 \\ 1 & 0 \end{bmatrix}\lambda + \begin{bmatrix} 1 & 0 \\ 0 & 0 \end{bmatrix}\beta\right)y_2 = 0,$$

and Eq. 14 with the definitions $\alpha = \chi^2, \beta = \Upsilon^2, \gamma = \Upsilon\chi$:

$$((M_0 + G_0)\alpha + G_2\beta + G_1\gamma - D_0\chi - K_0)x = 0,$$
$$((\overline{M}_0 + \overline{G}_0)\alpha + \overline{G}_2\beta + \overline{G}_1\gamma - \overline{D}_0\chi - \overline{K}_0)\overline{x} = 0,$$
$$\left(\begin{bmatrix} 1 & 0 \\ 0 & 0 \end{bmatrix}\alpha + \begin{bmatrix} 0 & 1 \\ 1 & 0 \end{bmatrix}\chi + \begin{bmatrix} 0 & 0 \\ 0 & 1 \end{bmatrix}\right)y_1 = 0, \qquad (30)$$
$$\left(\begin{bmatrix} 0 & 1 \\ 0 & 0 \end{bmatrix}\alpha + \begin{bmatrix} 0 & 0 \\ 1 & 0 \end{bmatrix}\beta + \begin{bmatrix} 1 & 0 \\ 0 & 1 \end{bmatrix}\gamma\right)y_2 = 0,$$

These systems can be solved via the operator determinant method, in a more general form than that presented in Section 4.2.

**5.2. The general operator determinant method**

The general form of a nonhomogeneous MEP is

$$W_i(\boldsymbol{\eta}) = A_{i0}x_i + \sum_{j=1}^{N} \eta_j A_{ij} x_i, \qquad i = 1, \ldots, n \qquad (31)$$

where $\boldsymbol{\eta}$ is the vector of eigenvalues ($\eta_j$ being the individual eigenvalues), $A_{ij}$ are the coefficient matrices, which can be complex and of different sizes for each equation, and $x_i$ are the eigenvectors. Eq. 31 can be visualized as a non-square array of matrices:

$$\begin{bmatrix} A_{10} & A_{11} & A_{12} & \cdots & A_{1N} \\ A_{20} & A_{21} & A_{22} & \cdots & A_{2N} \\ \vdots & \vdots & \vdots & \ddots & \vdots \\ A_{N0} & A_{N1} & A_{N2} & \cdots & A_{NN} \end{bmatrix}. \qquad (32)$$

In a process analogous to that presented in Section 4.4, we can construct a series of operator determinants for this system. There are $N + 1$ such operator determinants ($\Delta_0$ through to $\Delta_N$), of size $n^N$ for constant coefficient size $n$. They correspond to taking determinants of Eq. 32, with certain columns removed or inserted, and with the normal scalar multiplication operation replaced by a Kronecker product between matrices. Definitions and derivations of these determinants may be found in [40–44], and one software implementation in [39]. In the two-parameter case this analysis collapses into that of Section 4.2.

These general operator determinants allow us to compute the eigenvalues of the Eq. 31 via generalized eigenvalue problems. Providing $\Delta_0$ is nonsingular, it holds that

$$\Delta_i \mathbf{x} = \eta_i \Delta_0 \mathbf{x}. \tag{33}$$

In this way we are able to solve the quasi-linearized systems given in Section 5.1 (we will discuss singularity in Section 5.4). However, note that already by using this quasi-linearization process we have made the operator determinants smaller than they were with standard linearization. Using Eq. 12, the operator determinants are square and of size $2n^2$, and using Eq. 13 or 14 they are of size $4n^2$. With standard linearization they are of size $4n^2$ and $9n^2$, respectively.

### 5.3. Computing operator determinants

In the two-parameter case, the computation of the operator determinants is trivial. However in the case of larger systems we must find some other approach. In the general case, the operator determinants may be computed by modifying the Leibniz formula for the determinant [40]:

$$\Delta(\mathrm{M}) = \sum_{\mathbf{s} \in S_N} \mathrm{sgn}(\mathbf{s}) \bigotimes_{i=1}^{N} \mathrm{M}_{i,\mathbf{s}_i}, \tag{34}$$

where $S_N$ is the set of permutations of the set $\{1, \dots, N\}$, $\mathrm{sgn}(\mathbf{s})$ is the sign of the permutation vector $\mathbf{s} \in S_N$. M is the square matrix array (a modification of Eq. 32) corresponding to the operator determinant desired. The notation $\bigotimes_i^N$ denotes the repeated application of the Kronecker product. Note that the tensor determinant definition in [40] is slightly erroneous as the factor $(-1)^{\mathrm{sgn}(\sigma)}$ in their tensor determinant expressions should be either $\mathrm{sgn}(\sigma)$ or $(-1)^{N(\sigma)}$, where $N(\sigma)$ is the number of inversions in $\sigma$. Alternatively, Muhič and Plestenjak [39] devised an operator determinant Laplace expansion [45] that is able to compute the operator determinants of a system of arbitrary size, via a process of recursion:

$$\Delta(\mathrm{M}) = \sum_{i=1}^{N} (-1)^{i+1} \mathrm{M}_{i1} \otimes m_{i1}, \tag{35}$$

where $m_{i1}$ denotes the minor entry corresponding to $\mathrm{M}_{i1}$. The summation in Eq. 35 follows the first column of M, though, of course, many other summation paths could be used. It should be noted that the use of the Laplace or Leibniz methods for computing the determinant of an ordinary matrix have very high non-polynomial computational complexity costs – $\mathcal{O}(n!)$ and $\mathcal{O}(n!\,n)$, respectively [46,47].

### 5.4. Singularity

One major advantage of the direct solver based on quasi-linearization is that it generates linearized problems that are not always singular – when the coefficient matrices are of full rank, Eq. 28-30 are in general nonsingular. However in our case the coefficient $G_2$ is singular, and so it happens that all these three equations are singular anyway. The theory of the compression process noted in Section 4.3 has never been extended to the general multiparameter case, and so we have no rigorous method of solving singular MEPs with $N > 2$. However, we do have a practical method of doing so. Numerical experiments (some of which are detailed in Section 6) indicate that, although the compression algorithm takes only $\Delta_0$, $\Delta_1$ and $\Delta_2$ as an input, it will successfully compress these three operator determinants for the general multiparameter problem (ignoring all the others). In the case of Eq. 28-30, with these three operator determinants we can solve for all the eigenvalue parameters of the system, irrespective of the order in which we arrange the eigenvalues.

We should note that there is no justification for this procedure other than the experimental evidence that the algorithm works – evidence confirmed also by the nonrigourous $N > 2$ compression process utilized in [39]. The common regular part relationship proved by [29] applies only to two-parameter eigenvalue problems arising from linearization, and indeed it is not clear how the concept of the common regular part would extend to an $N > 2$ problem with more than two pencils. That the two-parameter algorithm does work for the first three operator determinants of an $N > 2$ problem implies that the algorithm will generalize – it is a question of working out what this generalization is. This is an interesting area for further research. For the purposes of this paper, however, we will use the compression algorithm for the quasi-linearized $\Delta_0$, $\Delta_1$ and $\Delta_2$ without proof.

## 6. Numerical experiments

### 6.1. Undamped model

We are now in a position to compute the flutter points of the models we introduced in Section 3. We have devised two direct solvers to do so: a direct solver with linearization and a direct solver with quasi-linearization. Consider first the undamped section model (Eq. 12), with parameter values as per Table 1. To validate our direct solution methods, we first produce a modal damping plot (Figure 2). Three points of neutral stability may be observed $\tau = \pm 1.00$ and $\tau = 0$. The point $\tau_F = 1.00$ is the only physical flutter point. We can link the modal damping curve for this flutter point with the lower modal frequency path in the $\text{Re}(\chi)$ plot (this requires data not observed directly on Figure 2). We thus can estimate that $\chi_F = 1.32$ rad/s ($\Lambda_F = 0.57\text{s}^2/\text{rad}^2$). The point at $\tau = -1.00$ along the same mode corresponds to a nonphysical flutter event occurring at negative airspeed. Finally, at $\tau = 0$ both modes have neutral stability (at $\chi = 0.548$ rad/s and $\chi = 1.426$ rad/s or $\Lambda = 3.33 \text{ s}^2/\text{rad}^2$ and $\Lambda = 0.492\text{s}^2/\text{rad}^2$). This, however, merely represents the fact that the structure is undamped at zero airspeed due to the lack of any structural damping in the model. The divergence point of this system does not appear on this plot, as it occurs at infinite $\tau$ and $\Lambda$. However, the computation of divergence points does not require multiparameter methods anyway, as if the frequency $\chi$ is assumed to be zero in the governing equation (e.g. Eq. 9) then the problem becomes a single parameter eigenvalue problem in the airspeed parameter (e.g. $\Upsilon$). This problem can then be solved with single-parameter solvers.

We then compute the flutter points of the system via our two direct solution methods, yielding flutter points identical to those detailed above for the presented accuracy. With the solver based on linearization, the uncompressed operator determinants are of size 16 and the compressed ones of size 4. With the solver based on quasi-linearization the size of the uncompressed determinants can be reduced significantly – down to 8 – while producing compressed determinants of the same size. The quasi-linearization method thus reduces the time required for the compression process, as the finite-regular part extraction algorithm works by successively increasing the rank of the system (not all at once). However, for this problem the computation time is too small for any meaningful assessment: we will investigate computation time more fully in Section 6.3. Figure 3 shows the system flutter points superimposed on a contour plot [24]. As can be seen, there is an exact agreement between the direct solutions and the contour plot solutions (the intersections of $\text{Re}(d) = 0$ and $\text{Im}(d) = 0$).

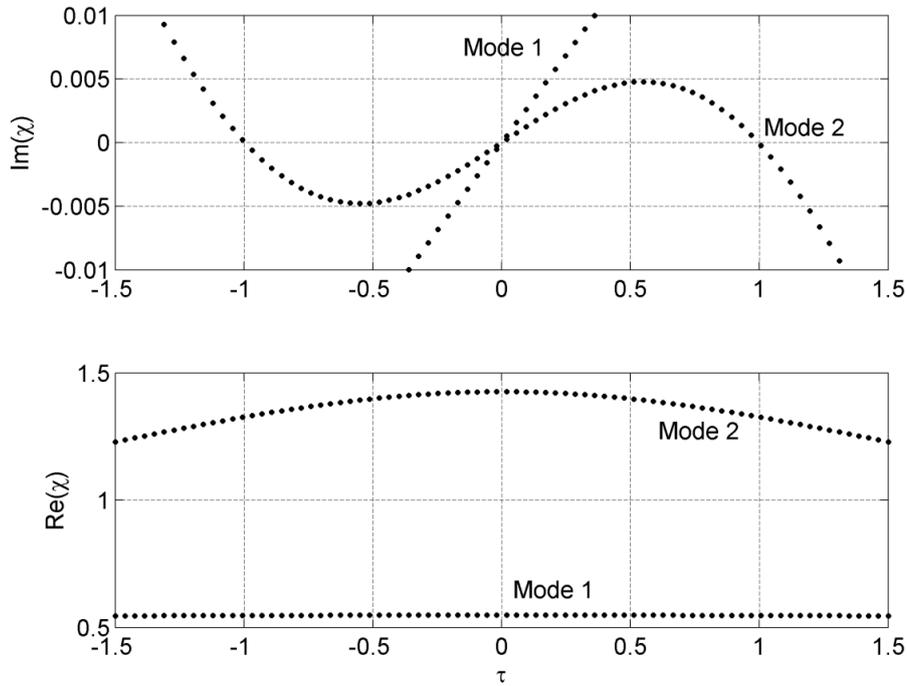

**Figure 2:** Modal damping plot for the undamped section model.

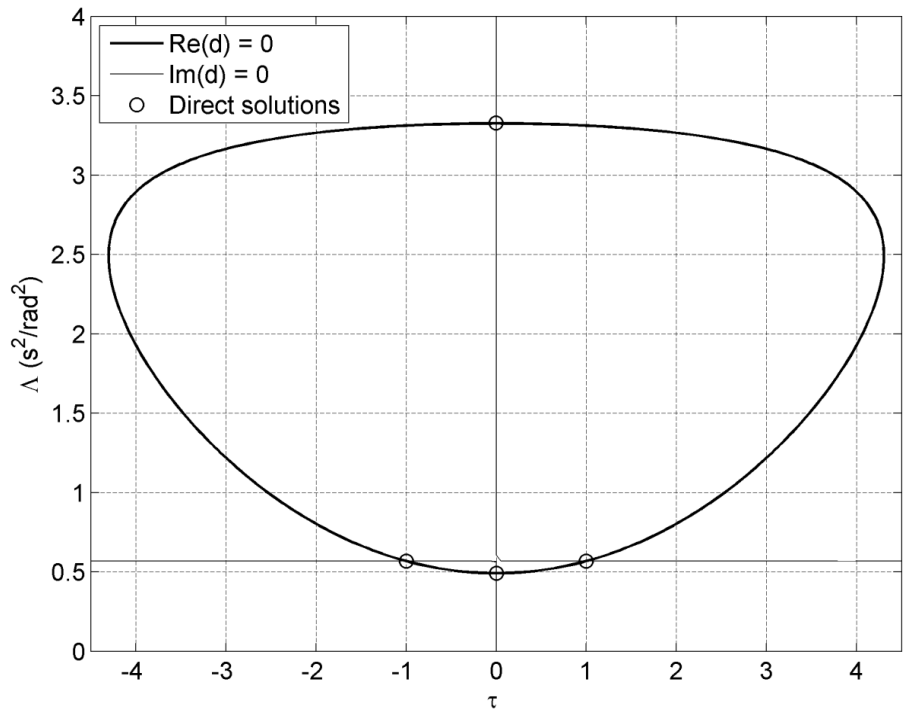

**Figure 3:** Contour plot for the undamped section model, showing identical solutions from the two direct solvers.

## 6.2. Damped model

We now simulate the damped system, in both forms (Eq. 13 and 14). Figure 4 shows a contour plot for the $\tau$-$\lambda$ form of this system (Eq. 16), with direct solutions from both solution methods. There is a single physical flutter point at $\tau_F = 1.650$, $\lambda_F = 0.8336$ s/rad, and a nonphysical flutter point at a small negative $\tau$. The divergence point occurs at infinite $\tau$ and $\lambda$. As expected, the solutions from both direct solution methods are identical to each other and the solutions from the contour plot. Figure 5 shows a contour plot of the $\Upsilon$-$\chi$ form (Eq. 14) with direct solutions from both solution methods. This physical flutter point can be located at $\Upsilon_F = 1.98$ Hz, $\chi_F = 1.20$ rad/s and the divergence point at $\Upsilon_D = 3.99$ Hz. Again, the flutter points computed with direct solvers agree exactly with those seen on the contour plot, and the results from the system as a whole agree with those of the $\tau$-$\lambda$ form ($\lambda_F = 1/\chi_F = 0.83$ s/rad and $\tau_F = \Upsilon_F/\lambda_F = 1.65$).

## 6.3. Computation time

Given the high computational complexity of these direct solution methods – $\mathcal{O}(n^6)$ – we are interested in the maximum system size for which a direct solution is practical. We have already found that for $n = 2$ the computational effort required is tiny, and so at the very least these direct solvers are useful for small reduced-order models as might be used in a preliminary design analysis. This alone is of use, as the directness of these solvers makes them ideal for use in optimization routines or other applications in which a large number of computations must be performed with limited user guidance. To gain a better understanding of the computational complexity characteristics of our methods, we simulate a series of systems of increasing matrix coefficient size. We generate random complex-valued matrices for the polynomial coefficients ($M_0$, $G_0$, etc.), of size $n = 2^k$ with $k \in \{1, 2, ..., 5\}$. For robustness, we average the results for $k = 1$ over 50 random matrices, for $k = 2$ over 10 random matrices, for $k = 3$ over 5 random matrices; and for systems larger than this we generate only one matrix. Figure 6 shows the wall-clock solution time against system size for the direct solver with linearization, for a 64-bit Intel i7-4770 with 3.4 GHz processor and 16 GB RAM, running MATLAB R2014b. Note that the $\lambda$ solution time denotes the time required to compute the $\lambda$ components of the solution via a series of one-parameter eigenvalue problems. As can be seen, the compression process is the most expensive component of the algorithm, making up a constant fraction of about 65% of the total computation time over the entire range of $n$. The GEP solution time is initially completely negligible but becomes more significant as system size increases: by $n = 32$ it makes up 34% of the total computation time. The $\lambda$-solution and setup process are never of much significance.

We expect from computational complexity theory that the gradient of the GEP-solution time curve in log-log units will be approximately 6.0 (Section 4.2). Fitting a linear curve through this data yields a gradient of 5.46, with an $R^2$ of 0.994. However, the GEP solution-time curve is slightly concave, with a maximum gradient of 6.47. The algorithm can then be said to have complexity $\mathcal{O}(n^6)$ overall. Figure 7 shows the wall-clock solution time against system size for the direct solver with quasi-linearization. This simulation was run on the same Intel i7-4770 platform with MATLAB R2014b. The random coefficient matrices used are identical to those in Figure 6.And although now the quasi-linearized system in no longer singular, we still apply the compression algorithm to capture some of its overhead costs.. As can be seen, these costs are not large but are still more significant than the GEP and $\lambda$ solution times for some system sizes.

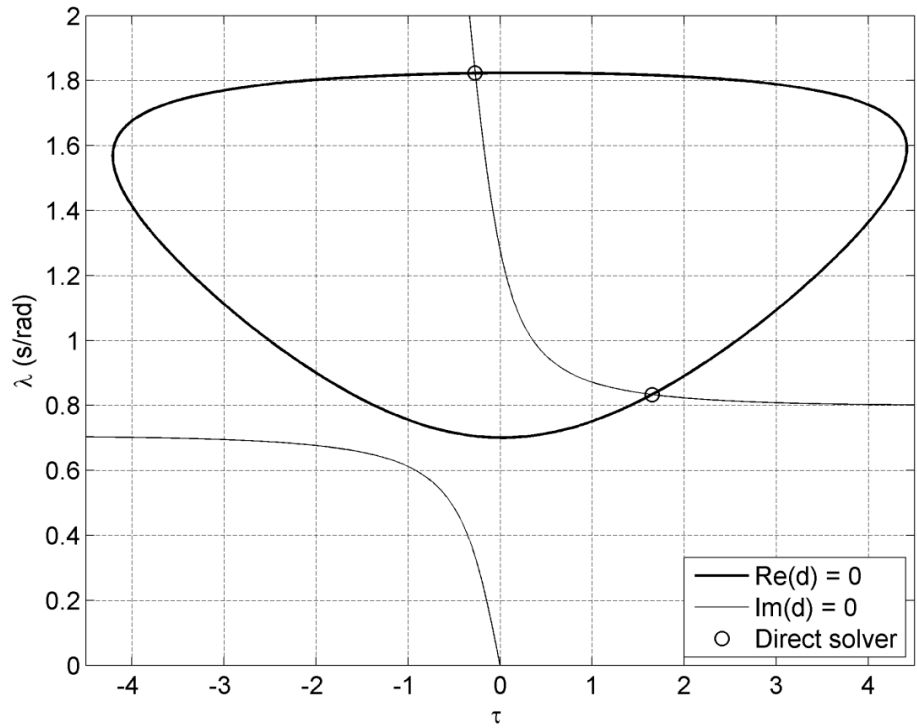

Figure 4: Contour plot for the damped section model ($\tau$-$\lambda$ form), showing identical solutions from the two direct solvers.

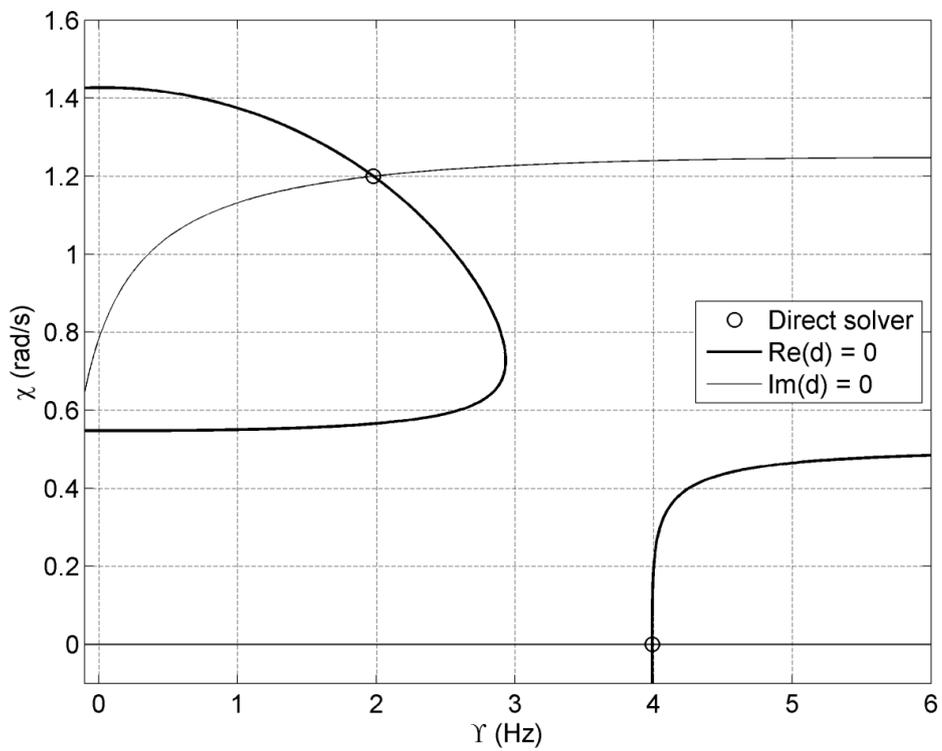

Figure 5: Contour plot for the damped section model ($\Upsilon$-$\chi$ form), showing identical solutions from the two direct solvers.

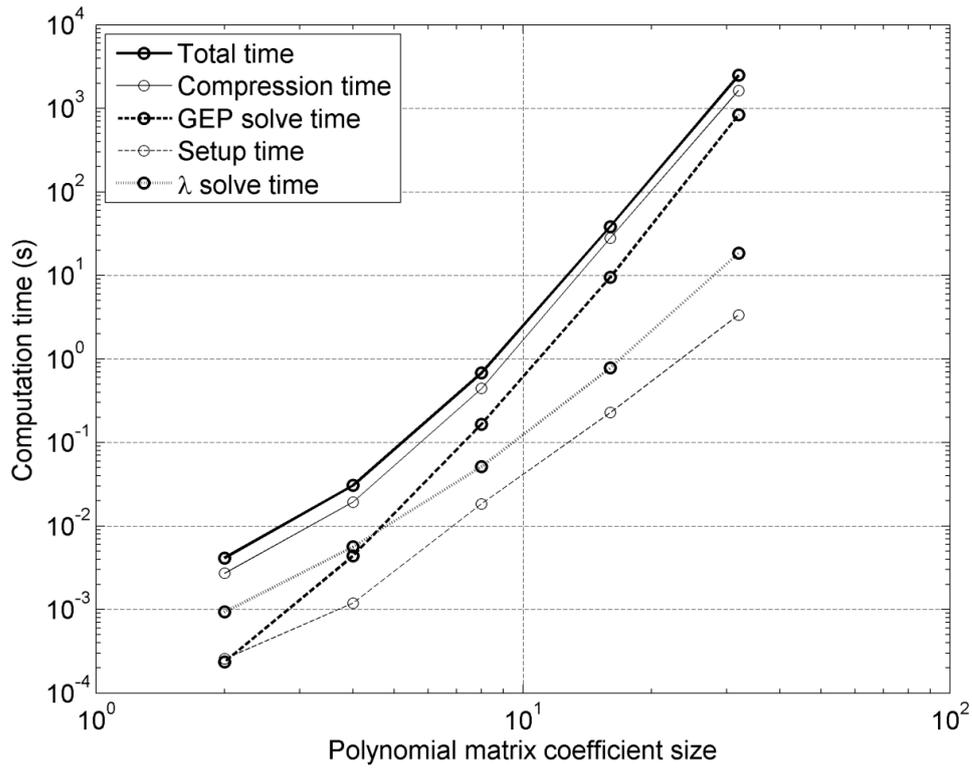

Figure 6: Wall-clock solution time against system size for the direct solver with linearization.

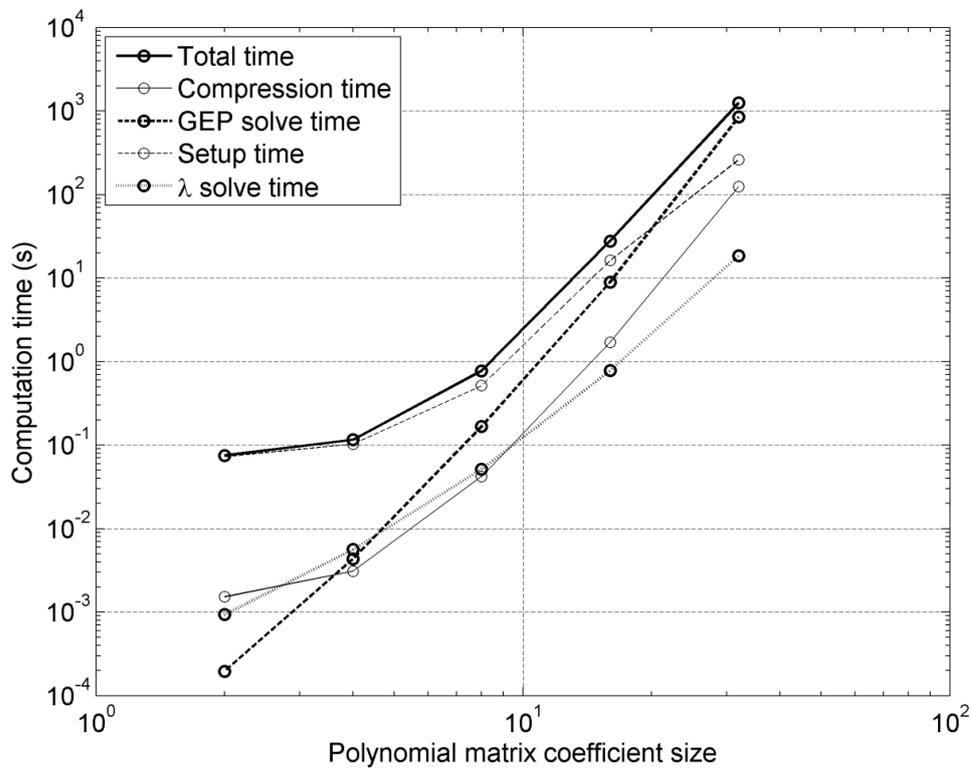

Figure 7: Wall-clock solution time against system size for the direct solver with quasi-linearization.

These GEP and $\lambda$ solution times are effectively identical to those of the direct solver with linearization, as the algorithm is solving a GEP of exactly the same size, yielding exactly the same eigenvalues. However, the most striking aspect of Figure 7 is the fact that the setup time occupies the majority of the required computation time right up to $n = 16$ (after which it is surpassed by the GEP solution time). The setup process involved using defining the system array (Eq. 32) in a cell array, and computing the necessary operator determinants of this array ($\Delta_0$ and $\Delta_2$) using the modified Leibniz formula (Eq. 34). On investigating these two procedures we find that it is the computation of the Kronecker products within the operator determinant which occupies the greatest time. However, despite this, the actual computational complexity of the algorithm as a whole is lower than that of the direct solver with linearization. GEP solution time is still approximately $\mathcal{O}(n^6)$, though the concavity that was noted in Figure 6 is still present.

Finally, Figure 8 shows a comparison of the total computation times for the two algorithms. As can be seen, below a system coefficient size of approximately 10, linearisation is faster than quasi-linearisation (at $n = 2$, over an order of magnitude faster). Above this coefficient size, quasi-linearisation is faster (at $n = 32$, twice as fast). To keep computation times below 10s, the system size must be below 11; to keep it below 1s, 8, and below 0.1s, 5. As it stands, the operator determinant method is not suitable for use with finite-element models or any other systems with a large number of degrees of freedom. It is, however, useful for the solution of simple reduced order models, e.g. in a preliminary design optimization.

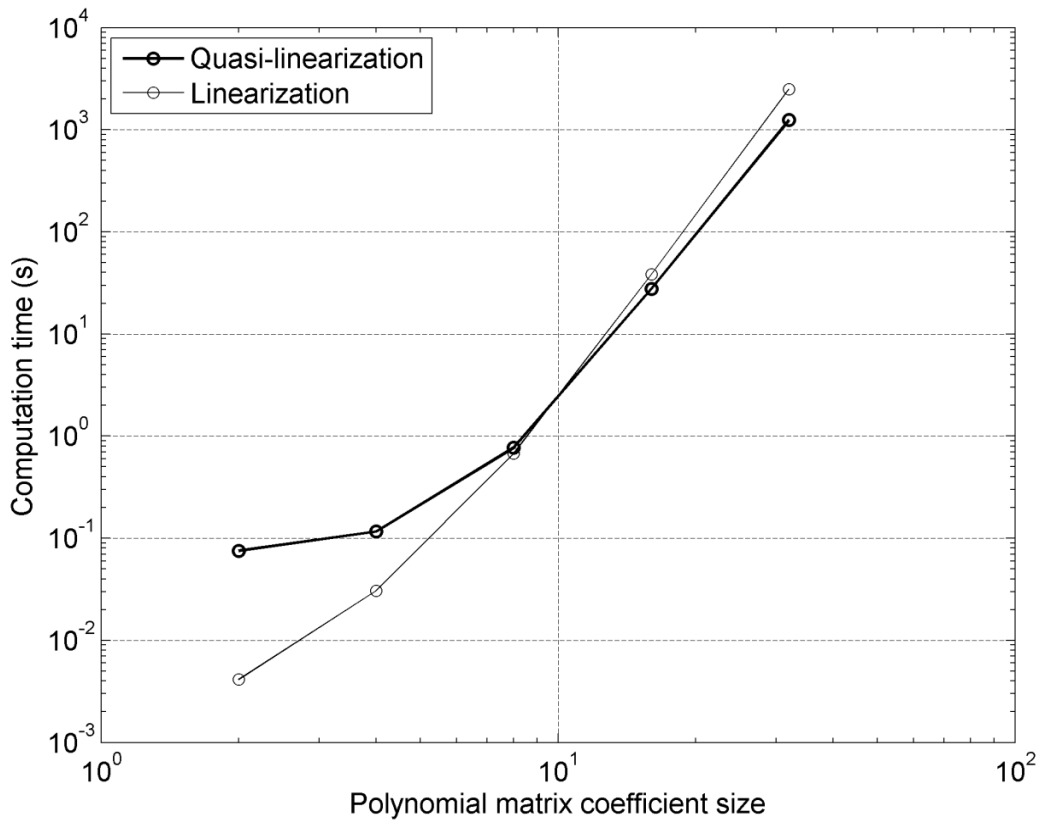

Figure 8: Total wall-clock solution time against system size for the two direct solver variants.

## 7. Discussion

### 7.1. Alternative solution methods

We have so far discussed the operator determinant method – a solution method which is direct but computationally expensive. However, two other classes of method are also available. The Sylvester-Arnoldi type methods are a set of closely-related algorithms that are only valid for two-parameter problems, but are fast and can handle large systems [48]. They still use operator determinants to reformulate the system into a set of generalized eigenvalue problems, but instead of solving this set of GEPs directly, the solution procedure is optimized based on the given knowledge that the operator determinants consist of a difference of two Kronecker products. For example, each step of a Krylov subspace procedure solution procedure for the GEP reduces to the solution of a Sylvester equation. Several related algorithms may be devised this way, and in some cases the computational complexity of the solution process can be reduced all the way down to $\mathcal{O}(n^3)$ [48]. However, this technique shows little potential for generalization to $N$-parameter systems, as it relies on being able to reformulate the operator determinant GEP into a simple and well-known matrix equation. As the operator determinant definition becomes more complex, the resulting matrix equation changes and efficient solvers may not be available. There is also no easy extension for singular problems.

Subspace methods for one-parameter eigenvalue problems are based around generating a series of linear spaces that eventually approximate one of the system's eigenspaces (the linear space of eigenvectors corresponding to a given eigenvalue). The Jacobi-Davidson and Rayleigh-Ritz methods are well-known one-parameter subspace methods, which can be generalized to apply to two-parameter systems [26,32,49–51], though this is not without difficulties [49]. These methods do not invoke the operator determinants, and show potential for generalization both to $N$-parameter and to polynomial systems. The Jacobi-Davidson is applicable to singular systems and has previously been tested on singular linearized aeroelastic stability problems [25]; but its performance was observed to be poorer than the operator determinant method. However the solution of singular MEPs has only been achieved very recently (2010 [29]), and so it is likely that the next few years will bring significant developments in this area.

### 7.2. Extensions to the concept

The core methodology presented in this paper can be extended to a wide variety of problems. Not only more complex two-parameter flutter models – considered in Pons [24] and Pons and Gutschmidt [25] – but also stability problems from a wide variety of fields, including systems with entirely different eigenvalue definitions. Indeed, any combination of model parameters in a stability model can indeed be treated as multiparameter eigenvalues; and scalar constraints on these parameters can cast as eigenvalue problems by introducing a scalar eigenvector [24]. For example, in aeroelastic model, given the location of a flutter point, we could solve for the sets of parameter values that could generate such a flutter point – allowing us to perform model identification based on flutter point information. This could pave the way for a least-squares approach for overconstrained multiparameter systems. Alternatively, we could introduce a flight altitude / air density / Mach number parameter and compute points on the aeroelastic flutter envelope of the aircraft. The multiparameter formulation provides a versatile way of analysing stability problems in any combination of parameters.

## 8. Conclusion

In this paper we have demonstrated and discussed the use of multiparameter solution techniques for the solution of aeroelastic and related stability problems. We have introduced the link between multiparameter spectral theory and stability analysis, and we showed how this link can be used to reformulate stability problems with a complex-valued stability metric and a pertinent environmental parameter into a two-parameter eigenvalue problem. We demonstrated that this allows the direct solution of stability problems that are linear or polynomial in these parameters, and we discussed aspects of the solution process, including the linearization and quasi-linearization of polynomial problems, general N-parameter problems, computational costs and approaches to problem singularity. We discuss extensions to these methods, including the generalization to more complex problems; and further applications, including parameter identification and flight envelope computation. The application of multiparameter methods to stability problems – in aeroelasticity and in other disciplines – has the potential to provide a wide variety of new methods for stability analysis.